\documentclass[smallextended]{svjour3}       
\smartqed  
\usepackage{graphicx}
\usepackage{color}
\usepackage{csvsimple}
 \journalname{Scientometrics}
\begin{document}

\title{Bibliometric analysis of topic structure in quantum computation and quantum algorithm research
\thanks{This work was supported by MEXT Quantum Leap Flagship Program (MEXT Q-LEAP) Grant Number JPMXS0120319794. }
}

\titlerunning{Bibliometric analysis of quantum conputation and quantum algorithms}        

\author{Tsubasa Ichikawa         
}


\institute{T. Ichikawa \at
              Center for Quantum Information and Quantum Biology, Osaka University, Osaka 560-0043, Japan \\
              \email{ichikawa@qiqb.osaka-u.ac.jp}           
}

\date{Dated: \today}

\maketitle

\begin{abstract}
We present a bibliometric analysis of the research papers on quantum computation and quantum algorithms published in 1985-2020.
We identify three distinct periods from the trend of the annual number of published papers,
and show the 20 top contributing countries in each period in terms of the number of publications and the number of total citations. 
The bibliographic coupling network of the publications in the latest period is characterized as a dense, small-world network with a small diameter, which contains 14 large communities, whose topics are the fabrication of qubits in various physical systems, studies of the quantum algorithms, and other related topics.
Quantum machine learning, one of the emerging topics in quantum computation, is found to be the fifth-largest independent community.
%
\keywords{bibliometrics \and bibliographic coupling \and Louvain method \and citation analysis \and quantum computation \and quantum algorithms}
\end{abstract}

\section{Introduction}
\label{intro}

Quantum computation (QC)  is an interdisciplinary research field, aiming at the implementation of quantum algorithms (QA), which utilize the principles of quantum mechanics \cite{NC2000}.
QC and QA have attracted much attention from researchers since the discovery of QA outperforming the corresponding classical algorithms. For example, Shor\rq s factorization algorithm \cite{Shor1994} performs prime factoring exponentially faster than the classical algorithms known so far.

Moreover, the last five years have witnessed the growing attention to QC and QA:
the U.S. national quantum initiative \cite{Monroe2019,Raymer2019} came into force in 2018, and other countries also started governmental investigations \cite{Riedel2019,Yamamoto2019,Sussman2019,Robertson2019} to make an ecosystem of quantum information science and technology (QIST).

With the growing interest in the QIST, there are several works to perform quantitative analyses on the state of the art of the QIST \cite{Tolcheev2018,Olijnyk2018,Farinhold2019,Chen2019,Scheidsteger2021,Dhawan2021,Seskir2021,Brito2021}.
These works clarified that i) the number of publications in the QIST is rapidly growing, ii) the leading countries of the QIST research are the US and China, and iii) Institutions in Europe have more collaborative works than those in the other countries.
In addition to these, more specific analyses have been conducted in quantum cryptography \cite{Olijnyk2018} and quantum machine learning \cite{Dhawan2021}.
The collaboration network in quantum information authors is found to be a small-world network \cite{Brito2021}.

This paper supports these findings and appends new insights from the viewpoint of the complex network analysis \cite{Barabasi2016}.
More precisely, we clarify the long-term trends of the researches on QC and QA from 1985 to 2020 by bibliometric analysis and show that the bibliographic coupling network \cite{Kessler1963} of QC and QA papers published in 2014-2020 is a dense, small-world network with a small diameter, which has a clearly distinct topic structure with emerging topics.
Thus, this work implies that QC and QA researches are mutually related deeply, and growing rapidly with the new topics emerging.

The rest of the paper is organized as follows:
In Sec.~2, we give a brief account of the method employed here.
In Sec.~3, we show the results obtained in our analyses.
Basic statistics of the dataset and the community structure of the topics are presented.
Section 4 is devoted to our conclusion.

\section{Method}
\label{method}
\paragraph{Data collection.} 
We collected 14,450 papers from the Clarivate Web of Science (WoS) database on 26 March 2021. We applied a query to the following attributes of the WoS database: abstracts, titles, author keywords, and keyword plus. 
By referring to \cite{Farinhold2019}, we constructed the query such that
\begin{quote}
(\lq\lq quantum computer\rq\rq~\textsf{OR} \lq\lq quantum computing\rq\rq~\textsf{OR} \lq\lq quantum compute\rq\rq~\textsf{OR} \lq\lq quantum computers\rq\rq~\textsf{OR} \lq\lq quantum algorithm\rq\rq~\textsf{OR} \lq\lq quantum algorithms\rq\rq~\textsf{OR} \lq\lq quantum inspired\rq\rq) 

\textsf{NOT} 

(\lq\lq post-quantum\rq\rq~\textsf{OR} \lq\lq quantum resistant\rq\rq~\textsf{OR} keywords:cryptography \textsf{OR} \lq\lq quantum key\rq\rq~\textsf{OR} conscious*~\textsf{OR} brain~\textsf{OR} dream~\textsf{OR} QTAIM),
\end{quote}
which was designed to collect the papers with respect to QC and QA but not to quantum cryptography and its related topics. 
The difference between our query and that in \cite{Farinhold2019} is laid on the treatment of quantum-inspired algorithms, which has been recently drawing the interest of the quantum information community \cite{Arrazola2020}. Bearing this trend in mind, we used the keyword \lq\lq quantum inspired\rq\rq~to include the papers of this topic in the resultant dataset, whereas  \cite{Farinhold2019} used it to exclude such papers.

\paragraph{Preprocessing.} 
Out of the 14,450 papers, we extracted the 14,086 papers which were published until the end of 2020 and whose authors are not anonymous. In addition, we deleted one paper that caused an error in constructing the bibliographic coupling network \cite{Kessler1963}, but no qualitative influence to the descriptive statistics mentioned below. The above preprocessing resulted in the 14,085 papers to be analyzed in this paper.

\paragraph{Outline of Analyses.} 
We performed the following statistical analyses. Throughout our analyses, we used R and its libraries Bibliometrix \cite{Aria2017} and wordcloud2.
\begin{enumerate}
  \item We analyzed the annual number of the publications and annual transition of the average citation number per year and paper. From the behaviors of these, we divided the period 1985-2020 into three mutually exclusive subperiods, that is, 1985-2003, 2004-2013, and 2014-2020.
  \item For each subperiod, we obtained the descriptive statistics by country. The indices analyzed were: the number of the publications, total number of citations, and the average number of citations per paper.
  \item For the latest subperiod 2014-2020, we performed the community detection. We constructed the bibliographic coupling network, where the coupling strength between two papers is given by the Jaccard coefficient \cite{Jaccard1912} with respect to the similarity of the reference lists. 
 The clustering method we have employed was the Louvain method \cite{Brondel2008}, which is of modularity optimization algorithms and known to be fast and accurate. Size distribution of the communities obtained was analyzed.
  \item For the large communities having more than one hundred papers, we identified their topics through their WordClouds obtained from the word occurrence distribution over the abstracts, titles, author keywords, and keyword plus in the dataset. 
 To make the WordClouds, 24 stopwords were chosen from the frequently appearing words over all the papers in the total period 1985-2020. We list the stopwords in Table~\ref{sw}.
\end{enumerate}

\begin{table}[b]
\caption{List of the stopwords in this paper.}
\label{sw}       
\begin{tabular}{ccccc}
\hline\noalign{\smallskip}
algorithm & algorithms & also & based & can \\
computer & computing & gate & gates & information \\
number & paper & problem & proposed & quantum \\
results & show & state & states & system \\
systems & time & two & using & \\
\noalign{\smallskip}\hline
\end{tabular}
\end{table}

\section{Results}
\label{results}
\subsection{Overview: 1985-2020}
\label{overview}

\begin{figure}
 \includegraphics[width=0.5\textwidth]{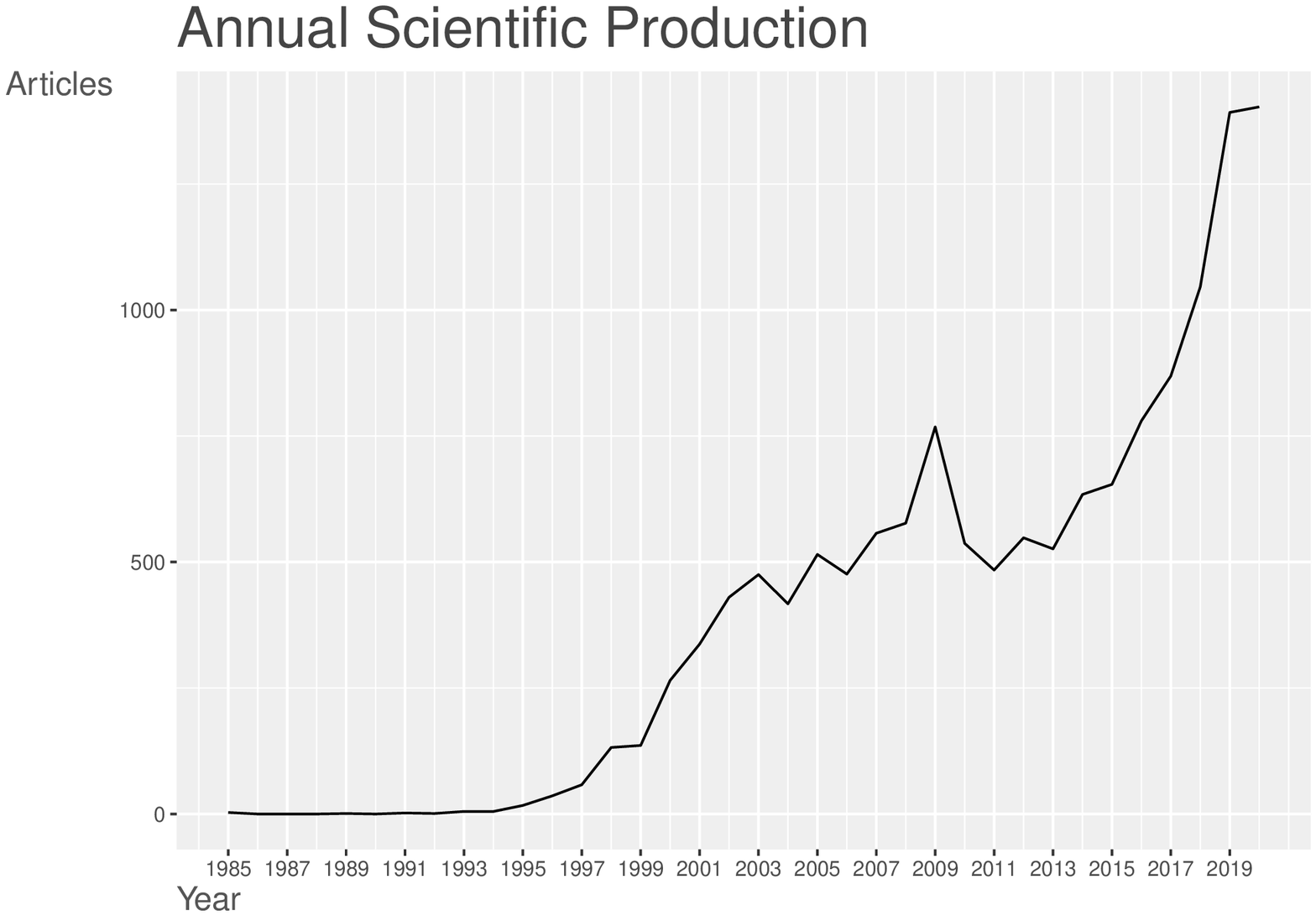}
 \includegraphics[width=0.5\textwidth]{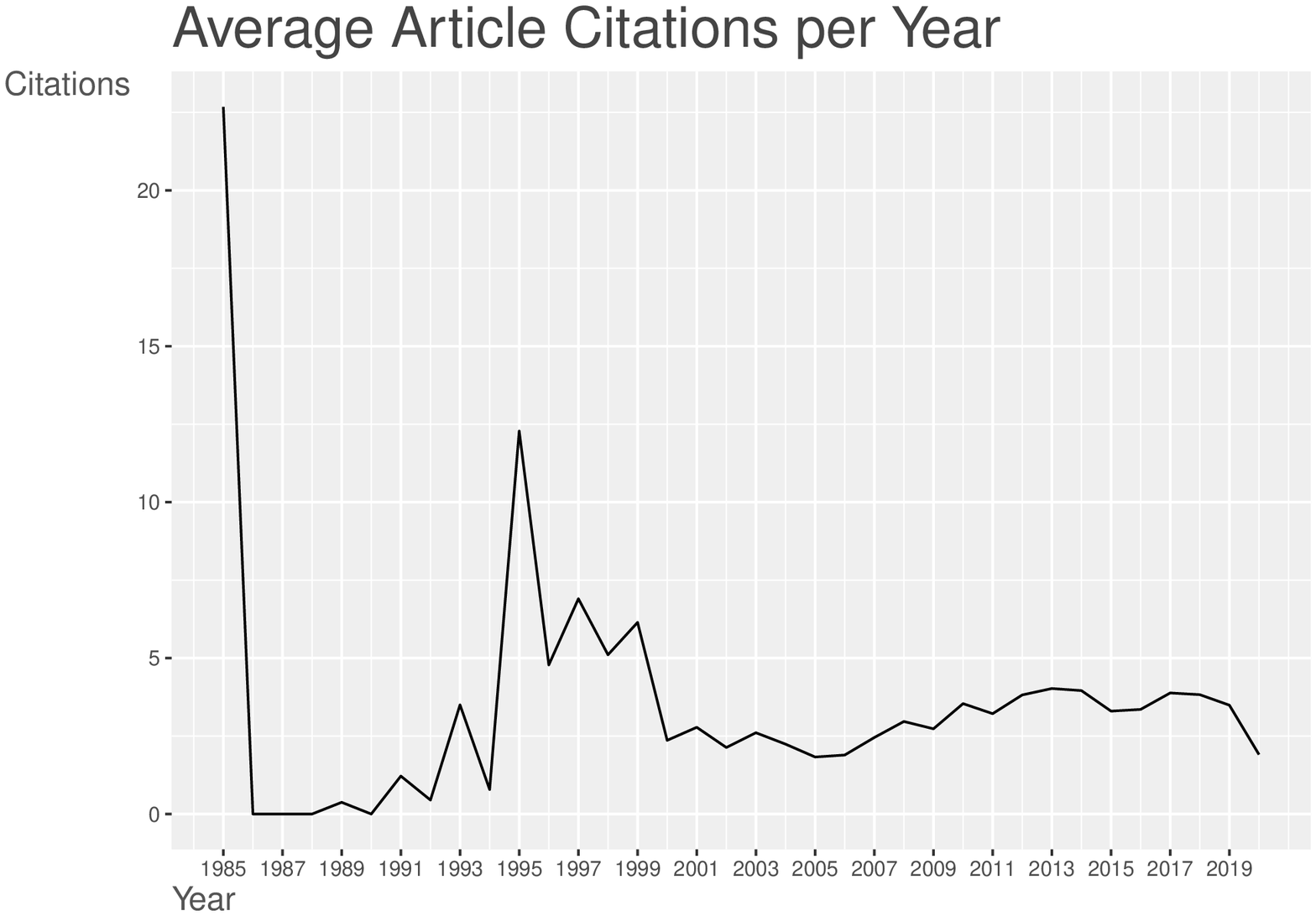}
\caption{(Left) The number of publications of QC and QA researches. (Right) The average citations of the papers per year.}
\label{fig:overview}       
\end{figure}

The left panel of Fig.~\ref{fig:overview} shows the annual shift of the number of publications in QC and QA researches. Whereas the recent publications, \cite{Fujii2015} for example, set the starting point of QC and QA research at the publication of a Feynman\rq s seminal paper \cite{Feynman1982} in 1982, it is not the case for the present study, since Feynman\rq s paper has no abstract and contains no query keywords in its title, and thereby out of our dataset. Instead, our dataset starts with the three papers published in 1985: the paper on universal quantum computation by Deutsch \cite{Deutsch1985}, that on the relation between reversible computation and quantum computation by Peres \cite{Peres1985}, and a brief explanation on \cite{Deutsch1985} by Maddox \cite{Maddox1985}, who was an editor of Nature at that time.

The number of publications had been almost negligible until 1993, but started to increase from 1994, which is the year that Shor\rq s factoring algorithm was put forward \cite{Shor1994}.
In 1985-2020, we can observe three distinct subperiods, 1985-2003, 2004-2013, and 2014-2020, from the behavior of the inclinations of the line graph.

Two relevant papers lie on the boundaries of these subperiods. One is a Kitaev\rq s paper published in 2003 \cite{Kitaev2003}, which proposed a quantum error-correcting code called the toric code. The toric code is distinguished from the other quantum error-correcting codes up until then, not only because its descendants, called surface codes, achieve high threshold values of error rate allowed to perform fault-tolerant quantum computation \cite{Fujii2015}, but also because it led to much literature on the investigation of topological matter. 
The other is a paper of the Martinis group in 2014 \cite{Barends2014}, which provided an experimental demonstration of the quantum elementary gates working with the error rate below the threshold.

The right panel of Fig.~\ref{fig:overview} shows the annual shift of the average article citations per year. The singular value in 1985 is because our dataset contains three papers in 1985, out of which \cite{Deutsch1985} and \cite{Peres1985} have been cited frequently (2048 citations for \cite{Deutsch1985} and 397 for \cite{Peres1985}). 
Putting this singular behavior in 1985 aside, we observe that relatively many highly cited papers had been published during 1995-1999, whence the average article citations per year has been almost stable. This implies that on the basis of the fundamental works performed in 1995-1999, QC and QA community has been grown steadily and established a stable citation network. Indeed, we can find the several papers in 1995-1999 which became the basic materials of the standard textbook of QC and QA community \cite{NC2000}: Shor\rq s 9 qubit code \cite{Shor1995}, Steane\rq s code \cite{Steane1996}, and Grover\rq s search algorithm \cite{Grover1998}, to name a few.

On the basis of the above observations, we may set three subperiods: 1985-2003, 2004-2013, and 2014-2020. The first subperiod 1985-2003 is the founding period, which began with Deutsch\rq s paper, went through the rapid increase of the publications, and ended up with the official acceptance of the toric code.
In contrast, the second subperiod 2004-2013 had been almost stable both in the number of annual publications and average article citations per year.
This stability ended up with the experiments conducted by the Martinis group, 
whence the number of published papers has rapidly increased with the average article citations per year being stable.

\subsection{Major countries in three subperiods}
\label{country}


\begin{figure}
(a) Founding period: 1985-2003.\\
 \includegraphics[width=0.7\textwidth]{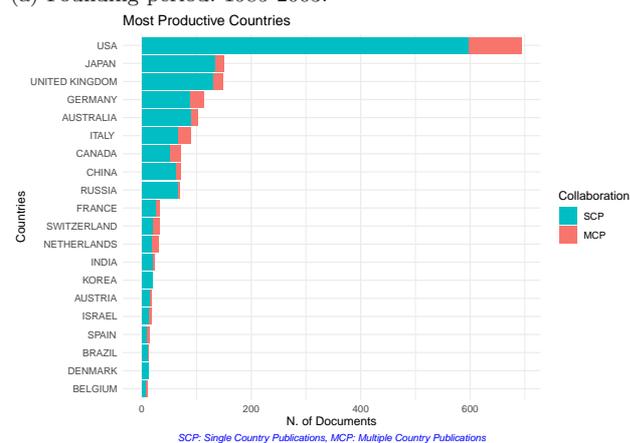}\\
(b) Stable period: 2004-2013.\\
 \includegraphics[width=0.7\textwidth]{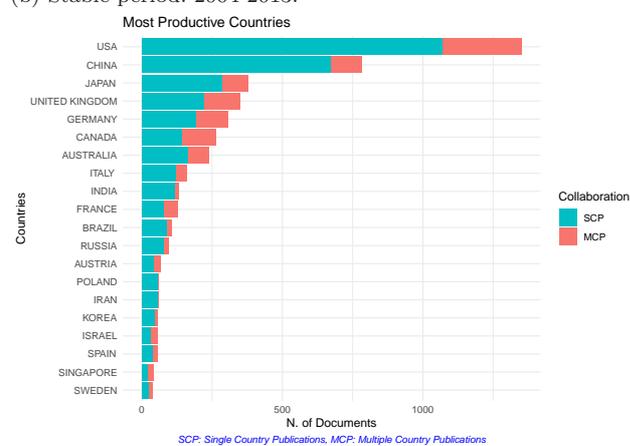}\\
(c) Growth period: 2014-2020.\\
 \includegraphics[width=0.7\textwidth]{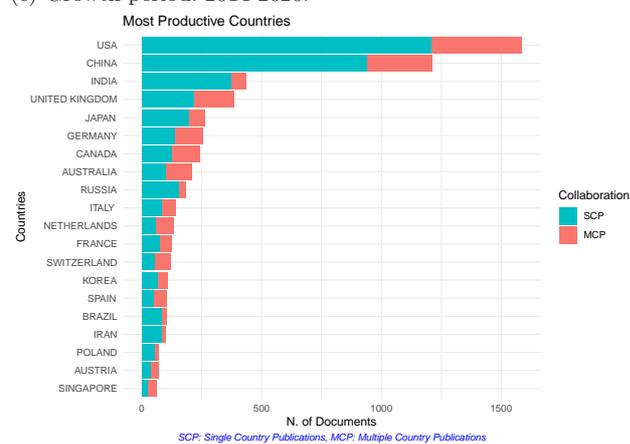}\\
\caption{Top 20 countries of the number of the papers in each subperiod. (a) 1985-2003, (b) 2004-2013, and (c) 2014-2020.}
\label{country_wise_pub}       
\end{figure}

Figure~\ref{country_wise_pub} shows the top twenty countries of the number of papers in each subperiod.
The US has the largest number of papers in each subperiod. On the other hand, the country with the second-largest number of publications is different in each subperiod: Japan in 1985-2003, and China in 2004-2013 and 2014-2020. In addition, the gap between the number of publications of the US and that of the second-ranked country has been narrowing in the second and third subperiods.
The ratio of multiple country publications has been increasing in many top countries. However, there are exceptional countries where the ratio of multiple country publications has been stable, such as Japan, Russia, and India.

\begin{table}
\caption{Top 20 countries of the total citations of the papers and their average article scitations in each subperiod. (a) 1985-2003, (b) 2004-2013, and (c) 2014-2020.}
\label{countries_table}       
(a) Founding period: 1985-2003.\\
\small
\begin{tabular}{c|rr}\hline%
Country      & Total Citations & Average Article Citations \\
\hline\hline
USA            & 68007 &  97.99 \\
UNITED KINGDOM & 12456 &  84.16 \\
GERMANY        &  7904 &  69.33 \\
AUSTRALIA      &  5556 &  54.47 \\
JAPAN          &  4929 &  32.86 \\
SWITZERLAND    &  4927 & 153.97 \\
CANADA         &  3214 &  44.64 \\
NETHERLANDS    &  3156 & 101.81 \\
AUSTRIA        &  2460 & 136.67 \\
ITALY          &  2203 &  24.75 \\
CHINA          &  1800 &  25.00 \\
RUSSIA         &  1575 &  22.83 \\
KOREA          &  1467 &  73.35 \\
SPAIN          &  1272 &  97.85 \\
ISRAEL         &  1028 &  57.11 \\
BELGIUM        &   998 &  90.73 \\
DENMARK        &   926 &  77.17 \\
FRANCE         &   879 &  26.64 \\
INDIA          &   528 &  22.00 \\
MEXICO         &   387 &  64.50 \\
 \hline
\end{tabular}\\

(b) Stable period: 2004-2013.\\
\begin{tabular}{c|rr}\hline%
Country      & Total Citations & Average Article Citations \\
\hline\hline
USA             & 63924 &  47.32 \\
UNITED KINGDOM  & 16037 &  45.95 \\
GERMANY         & 15063 &  49.07 \\
CANADA          & 13008 &  49.46 \\
CHINA           & 12488 &  15.99 \\
AUSTRALIA       & 12041 &  50.59 \\
JAPAN           &  6664 &  17.58 \\
AUSTRIA         &  6313 &  92.84 \\
NETHERLANDS     &  5568 & 154.67 \\
ISRAEL          &  4598 &  82.11 \\
FRANCE          &  4525 &  35.35 \\
ITALY           &  2789 &  17.32 \\
SPAIN           &  2594 &  46.32 \\
BRAZIL          &  2156 &  20.73 \\
DENMARK         &  1985 &  56.71 \\
SWITZERLAND     &  1871 &  66.82 \\
KOREA           &  1371 &  24.05 \\
RUSSIA          &  1262 &  12.88 \\
INDIA           &  1097 &   8.44 \\
SINGAPORE       &   631 &  15.39 \\
 \hline
\end{tabular}\\

(c) Growth period: 2014-2020.\\
\begin{tabular}{c|rr}\hline%
Country      & Total Citations & Average Article Citations \\
\hline\hline
USA             & 26466 & 16.70 \\
CHINA           & 11481 &  9.48 \\
UNITED KINGDOM  &  6222 & 16.08 \\
AUSTRALIA       &  4563 & 21.94 \\
GERMANY         &  4015 & 15.75 \\
JAPAN           &  3862 & 14.68 \\
NETHERLANDS     &  3026 & 22.42 \\
CANADA          &  3015 & 12.51 \\
INDIA           &  1798 &  4.12 \\
FRANCE          &  1733 & 13.75 \\
AUSTRIA         &  1729 & 24.70 \\
SWITZERLAND     &  1717 & 14.19 \\
ITALY           &  1488 & 10.33 \\
SPAIN           &  1357 & 13.05 \\
DENMARK         &  1339 & 38.26 \\
RUSSIA          &   905 &  4.95 \\
SOUTH AFRICA    &   783 & 27.00 \\
KOREA           &   580 &  5.23 \\
SWEDEN          &   550 & 14.47 \\
IRAN            &   497 &  4.92 \\
 \hline
\end{tabular}
\end{table}

Let us turn to Table~\ref{countries_table}, which shows the top twenty countries of the total citations of the papers and their average article citations in each subperiod. The ranking of total citations shows a similar trend to that of the total number of papers. This is a natural behavior since the total number of citations increases as the total number of papers increases. On the other hand, small countries in the population such as Austria, Denmark, and Switzerland, which are the countries of the founders of quantum mechanics, are highly ranked in average article citations. In particular, it is notable that South Africa has been ranked since 2014. This is because quantum machine learning  \cite{Wittek2014,Schuld2018} has been a promising research field in Noisy Intermediate-Scale Quantum (NISQ) computer research \cite{Preskill2018} since around 2014, and a strong research group in quantum machine learning is in South Africa.

The above observations show that although the US has been the leading country in QC research, the gap between the US and other countries is narrowing, and the internationalization of research is progressing. Besides, with the development of the NISQ computer researches, the influence of emerging countries is growing.


\subsection{Network statistics and community structure in 2014-2020}
\label{community}

\begin{table}[b]
\caption{Descriptive statistics of the bibliographic coupling network of the papers published in 2014-2020. The size is the number of the vertices $|V|$, and density is defined by $|E|/|V|^2$, where $|E|$ being the number of the edges. The square root density, denoted by $d$ in \cite{Melancon2006}, is defined by $\sqrt{|E|/|V|^2}$.}
\label{stats_net}       
\begin{tabular}{c|r}
\hline
Measure & Value  \\
\hline\hline
Size & 6777  \\
Average degree & 317  \\
Density & 0.091  \\
Square root density & 0.301  \\
Average path length & 2.056  \\
$\ln|V|/\ln k$ & 1.532  \\
Diameter & 7  \\
\hline
\end{tabular}
\end{table}

\begin{figure}
 \includegraphics[width=0.5\textwidth]{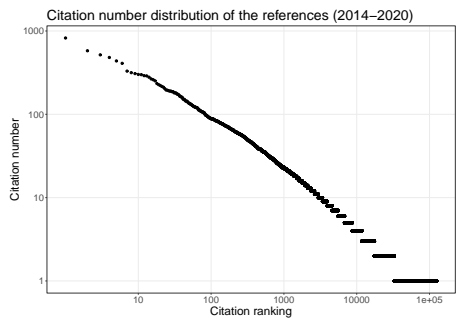}
\caption{Citation number-rank distribution of the references of the papers published in 2014-2020.}
\label{fig:degree_rank}       
\end{figure}

Let us hereafter focus on the latest subperiod 2014-2020.
As stated above, this term witnessed the emergence of new topics, which allow us to expect a rich topic structure.
To discuss this, we deal with the descriptive statistics of the bibliographic network of the papers published in 2014-2020 and the size-rank distribution of the community obtained by the Louvain method.

Table \ref{stats_net} shows the descriptive statistics of the basic measures of the bibliographic coupling network.
The square root of the network density is 0.301, which is higher than those of many real-world datasets analyzed in \cite{Melancon2006}.
This shows that the network considered here is relatively dense, suggesting that the papers of QC and QA share the basic literature often to be commonly cited. Indeed, the citation number distribution of the references in the papers published in 2014-2020 obeys a power law (Fig.~\ref{fig:degree_rank}).

The average path length of the small-world networks is approximated by $\ln |V|/\ln k$, where $k$ is the average degree \cite{Barabasi2016}.
The bibliographic network considered here has an average path length of 2.056, which is close to the value expected for the small-world network $\ln |V|/\ln k\approx1.532$.
This suggests that the bibliographic network is small-world property.
In addition to this, the bibliographic coupling network has a diameter of $7$. This is larger than that expected for the random network $\ln |V|/\ln k\approx1.532$ \cite{Barabasi2016}, and smaller than that for the high-energy physics citation network, whose diameter is 12 \cite{Gehrke2003,Leskovec2005}.
Summing up, the bibliographic network is a dense, small-world network with a small diameter.

\begin{figure}
 \includegraphics[width=0.5\textwidth]{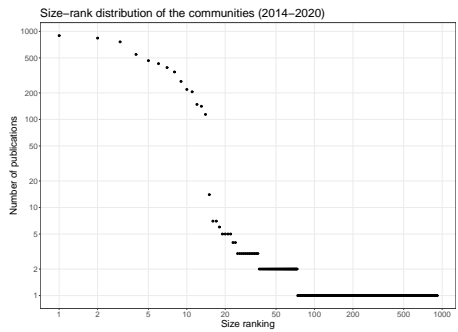}
\caption{The size-rank distribution of the communities obtained by the Louvain method from the bibliographic coupling network 2014-2020.}
\label{fig:size_rank}       
\end{figure}

Applying the Louvain method to the bibliographic coupling network, we obtained 911 communities, whose size-rank distribution is shown in Fig.~\ref{fig:size_rank}.
Out of the 911 communities, the fourteen largest communities consist of more than a hundred papers.
Besides, there exists a gap in the community size between the 14th and 15th communities:
whereas the 14th community consists of 114 papers, the 15th consists of 14 papers.
Such a gap is found in a community structure of an SNS network and reproduced by a connected nearest-neighbor model with random edge creation to non-nearest neighboring vertices \cite{Yuta2007}.

\subsection{Characters of major communities}
\label{character}

\begin{figure}
 \includegraphics[width=0.5\textwidth]{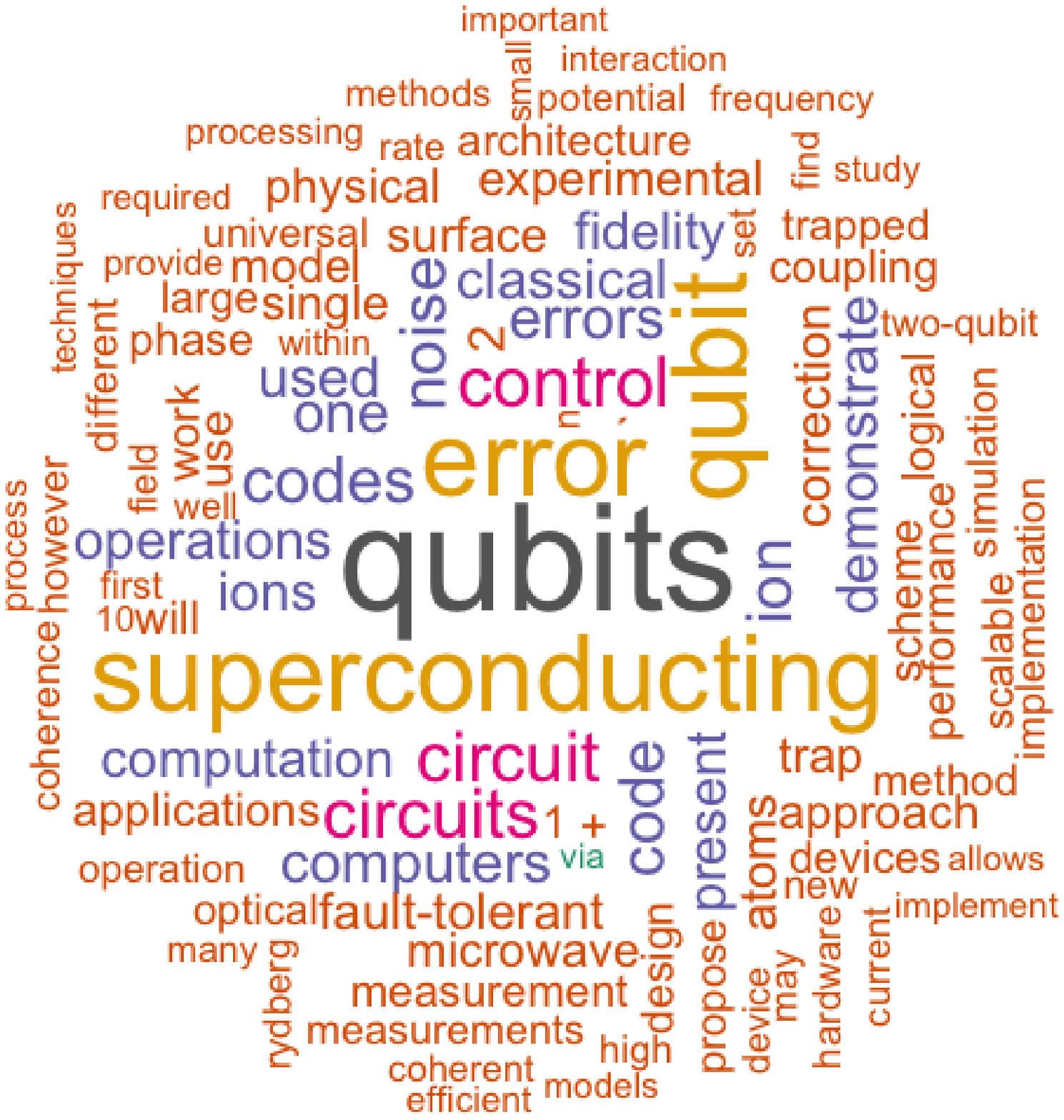}
\caption{The WordCloud of the largest community of the papers published in 2014-2020.}
\label{fig:wordcloud}       
\end{figure}

To find the subjects of the major communities mentioned above, we made the WordClouds of the words appearing in the abstracts of the papers, and identified the subjects.
For example, Fig.~\ref{fig:wordcloud} shows the WordCloud of the largest community, suggesting that its issues are of superconducting qubits and error tolerance.

Table \ref{summary_community} summarizes the statistics and subjects of the largest 14 communities, which contain 5767 papers (85.1\% of 6777 papers).
Three remarks are in order.
First, these 14 communities can be classified into three categories as follows:
\begin{enumerate}
\item Nine communities on QC and QA using quantum gates (No.~1, 2, 3, 4, 5, 6, 7, 11, 13), which contain 4670 papers.
\item One community on the adiabatic quantum computation and quantum annealing (No.~9) of 270 papers.
\item Four communities on the topics related to QC (No.~8, 10, 12, 14), which contain 827 papers.
\end{enumerate}
This shows that QC and QA using quantum gates attract much attention to the researchers and dominate the research field of QC.

\begin{table}[b]
\caption{Statistics of the major communities of the papers published in 2014-2020. The last row shows the statistics all the publications in this period.}
\begin{center}
\begin{tabular}{|r|r|r|r|l|}
\hline
No. & Publications & Ratio (\%) & Median & Subjects  \\
 &  &  & of citations &  \\
\hline\hline
1 & 895 & 13.2 & 4 & Fabrication and use of low-error qubits,\\
&&&& in particular, superconducting qubits. \\
2 & 838 & 12.4 & 3 & Photon quantum computing, boson sampling. \\
3 & 760 & 11.2 & 2 & Quantum algorithms, in particular, \\
&&&&application to image processing. \\
4 & 547 & 8.1 & 3 & Quantum simulations. \\
5 & 465 & 6.9 & 2 & Quantum machine learning. \\
6 & 430 & 6.3 & 5 & Topological matter and\\
&&&& topological quantum computation. \\
7 & 388 & 5.7 & 4 & Silicon qubits. \\
8 & 346 & 5.1 & 3 & Quantum inspired algorithms. \\
9 & 270 & 4.0 & 3 & Adiabatic quantum computing \\
&&&& and quantum annealing. \\
10 & 219 & 3.2 & 1 & Construction and implementation\\
&&&& of reversible logic gates. \\
11 & 206 & 3.0 & 5 & NV center qubits. \\
12 & 148 & 2.2 & 1 & Post-quantum cryptography. \\
13 & 141 & 2.1 & 9 & Molecular spin qubits. \\
14 & 114 & 1.7 & 3 & Quantum walk. \\
\hline
Total & 6777 & 100.0 & 2 & \\
\hline
\end{tabular}
\end{center}
\label{summary_community}
\end{table}%


Second, the communities concerning the fabrication of qubits (No.~1, 6, 7, 11, 13) have medians of citations higher than that of all the papers published in 2014-2020.
This suggests the relevance of creating the qubits in good quality.
Indeed, it was pointed out as DiVincenzo\rq s criteria \cite{DiVincenzo2000} that building a quantum computer needs scalable systems with well-characterized qubits on which we can perform qubit-wise initialization, gate operations, and measurements within their coherent time.

Third, quantum machine learning, which is a relatively new research area, is an independent community. 
This suggests the rapid growth of quantum machine learning research. Indeed, whereas the composite annual growth rate (CAGR) of the publication numbers from 2014 to 2020 is 14.2\%, the CAGR of quantum machine learning publications is 55.4\%.


\section{Conclusion}
\label{conc}

In this paper, we investigated the trends and topics in the research of QC and QA.
We have found that the number of publications is rapidly growing since 2014, and since then the emerging countries participated in QC research.
Besides, we identified the major 14 topics of QC research in the period 2014-2020. We observed not only the longstanding topics such as the fabrication of the qubits in good quality, but also the emerging topic such as quantum machine learning.
In addition, we have clarified that the papers of QC and QA comprise a dense, small-world network with a small diameter. 
Notably enough, the size-rank distribution of the communities (sets of the papers sharing the same topics) has a gap between the sizes of the 14 largest communities and the others.
These findings suggest that in QC and QA research, the topics are mutually related deeply and rapidly growing, with new topics emerging.

In closing, we mention that it could be of interest that one applies the method of the present analysis to other QIST research fields such as quantum communication or quantum metrology, which have different origins from QC and QA, and thereby their topic structures would be expected to differ from that of QC and QA bibliographic coupling network.
It is also of importance to investigate a generation mechanism of the gap in the community size distribution.
These remaining issues could uncover characteristic features of the QIST research communities.

%
%


%
%



\end{document}